\documentclass[twocolumn,twocolappendix]{aastex63}
\usepackage{xcolor}

\newcounter{daggerfootnote}

\newcommand{\jwst}{\textit{JWST}}
\newcommand{\hst}{\textit{HST}}

\definecolor{mycol}{rgb}{0,0,1}

\received{n/a}
\revised{n/a}
\accepted{n/a}

\submitjournal{\apjl}

\shorttitle{Spatially Resolved \hst-dark Galaxy}
\shortauthors{Kokorev et al.}

\usepackage{longtable}
\usepackage{lipsum}
\usepackage{amsmath}
\usepackage[normalem]{ulem}
\usepackage[para]{threeparttable}
\usepackage{enumitem}

\begin{document}

 \title{\textit{JWST} Insight Into a Lensed \textit{HST}-dark Galaxy and its Quiescent Companion at $z=2.58$}

\correspondingauthor{Vasily Kokorev}
\email{kokorev@astro.rug.nl}

\author[0000-0002-5588-9156]{Vasily Kokorev}
\affiliation{Kapteyn Astronomical Institute, University of Groningen, P.O. Box 800, 9700AV Groningen, The Netherlands}

\author[0000-0002-8412-7951]{Shuowen Jin}\altaffiliation{Marie Curie Fellow}
\affiliation{Cosmic Dawn Center (DAWN), Denmark}
\affiliation{DTU-Space, Technical University of Denmark, Elektrovej 327, DK2800 Kgs. Lyngby, Denmark}

\author[0000-0002-4872-2294]{Georgios E. Magdis}
\affiliation{Cosmic Dawn Center (DAWN), Denmark}
\affiliation{DTU-Space, Technical University of Denmark, Elektrovej 327, DK2800 Kgs. Lyngby, Denmark}
\affiliation{Niels Bohr Institute, University of Copenhagen, Jagtvej 128, 2100, Copenhagen N, Denmark}

\author[0000-0001-8183-1460]{Karina I. Caputi}
\affiliation{Kapteyn Astronomical Institute, University of Groningen, P.O. Box 800, 9700AV Groningen, The Netherlands}
\affiliation{Cosmic Dawn Center (DAWN), Denmark}

\author[0000-0001-6477-4011]{Francesco Valentino}
\affiliation{Cosmic Dawn Center (DAWN), Denmark}
\affiliation{Niels Bohr Institute, University of Copenhagen, Jagtvej 128, 2100, Copenhagen N, Denmark}
\affiliation{European Southern Observatory, Karl-Schwarzschild-Str. 2, D-85748 Garching bei Munchen, Germany}

\author[0000-0001-8460-1564]{Pratika Dayal}
\affiliation{Kapteyn Astronomical Institute, University of Groningen, P.O. Box 800, 9700AV Groningen, The Netherlands}

\author[0000-0002-6849-5375]{Maxime Trebitsch}
\affiliation{Kapteyn Astronomical Institute, University of Groningen, P.O. Box 800, 9700AV Groningen, The Netherlands}

\author[0000-0003-2680-005X]{Gabriel Brammer}
\affiliation{Cosmic Dawn Center (DAWN), Denmark}
\affiliation{Niels Bohr Institute, University of Copenhagen, Jagtvej 128, 2100, Copenhagen N, Denmark}

\author[0000-0001-7201-5066]{Seiji Fujimoto}
\affiliation{Department of Astronomy, The University of Texas at Austin, Austin, TX 78712, USA}
\affiliation{Cosmic Dawn Center (DAWN), Denmark}
\affiliation{Niels Bohr Institute, University of Copenhagen, Jagtvej 128, 2100, Copenhagen N, Denmark}

\author[0000-0002-8686-8737]{Franz Bauer}
\affil{Instituto de Astrof{\'{\i}}sica, Facultad de F{\'{i}}sica, Pontificia Universidad Cat{\'{o}}lica de Chile, Campus San Joaquín, Av. Vicuña Mackenna 4860, Macul Santiago, Chile, 7820436} 
\affil{Centro de Astroingenier{\'{\i}}a, Facultad de F{\'{i}}sica, Pontificia Universidad Cat{\'{o}}lica de Chile, Campus San Joaquín, Av. Vicuña Mackenna 4860, Macul Santiago, Chile, 7820436} 
\affil{Millennium Institute of Astrophysics, Nuncio Monse{\~{n}}or S{\'{o}}tero Sanz 100, Of 104, Providencia, Santiago, Chile} 

\author[0000-0001-8386-3546]{Edoardo Iani}
\affiliation{Kapteyn Astronomical Institute, University of Groningen, P.O. Box 800, 9700AV Groningen, The Netherlands}

\author[0000-0002-4052-2394 ]{Kotaro Kohno}
\affiliation{Institute of Astronomy, Graduate School of Science, The University of Tokyo, 2-21-1 Osawa, Mitaka, Tokyo 181-0015, Japan}
\affiliation{Research Center for the Early Universe, Graduate School of Science, The University of Tokyo, 7-3-1 Hongo, Bunkyo-ku, Tokyo 113-0033, Japan}

\author[0000-0001-7880-8841]{David Bl\'anquez Ses\'e}
\affiliation{Cosmic Dawn Center (DAWN), Denmark}
\affiliation{DTU-Space, Technical University of Denmark, Elektrovej 327, DK2800 Kgs. Lyngby, Denmark}

\author[0000-0002-4085-9165]{Carlos G\'omez-Guijarro}
\affiliation{Universit{\'e} Paris-Saclay, Universit{\'e} Paris Cit{\'e}, CEA, CNRS, AIM, 91191, Gif-sur-Yvette, France}

\author[0000-0002-5104-8245]{Pierluigi Rinaldi}
\affiliation{Kapteyn Astronomical Institute, University of Groningen, P.O. Box 800, 9700AV Groningen, The Netherlands}

\author[0000-0001-6066-4624]{Rafael Navarro-Carrera}
\affiliation{Kapteyn Astronomical Institute, University of Groningen, P.O. Box 800, 9700AV Groningen, The Netherlands}

\begin{abstract}
Using the novel \textit{JWST}/NIRCam observations in the Abell 2744 field, we present a first spatially resolved overview of an \textit{HST}-dark galaxy, spectroscopically confirmed at $z=2.58$ with magnification $\mu\approx1.9$. While being largely invisible at $\sim$1 $\mu$m with NIRCam, except for sparse clumpy sub-structures, the object is well-detected and resolved in the long-wavelength bands with a spiral shape clearly visible in F277W. By combining ancillary ALMA and $Herschel$ data, we infer that this object is an edge-on dusty spiral with an intrinsic stellar mass log$(M_*/M_\odot)\sim11.3$ and a dust-obscured SFR$\sim 300~M_\odot$~yr$^{-1}$. A massive quiescent galaxy (log$(M_*/M_\odot)\sim10.8$) with tidal features lies 2\farcs{0} away ($r$$\sim$9 kpc), at a consistent redshift as inferred by \textit{JWST} photometry, indicating a potential major merger. The dusty spiral lies on the main-sequence of star formation, and shows high dust attenuation in the optical ($3<A_{\rm V}<4.5$). In the far-infrared, its integrated dust SED is optically thick up to $\lambda_0 \sim 500$ $\mu$m, further supporting the extremely dusty nature. Spatially resolved analysis of the \textit{HST}-dark galaxy reveals a largely uniform $A_{\rm V}\sim 4$ area spanning $\sim$57 kpc$^2$, which spatially matches to the ALMA 1 mm continuum emission. Accounting for the surface brightness dimming and the depths of current \textit{JWST} surveys, unlensed analogs of the \textit{HST}-dark galaxy at $z>4$ would be only detectable in F356W and F444W in UNCOVER-like survey, and become totally \textit{JWST}-dark at $z\sim6$. This suggests that detecting highly attenuated galaxies in the Epoch of Reionization might be a challenging task for \textit{JWST}.

\end{abstract}


\section{Introduction} \label{sec:intro}
The last few decades of observations with the Hubble Space Telescope (\hst), in conjunction with an array of infrared (IR) and millimeter facilities, have revealed a population of dust obscured, optically faint, star-forming galaxies. Compared to massive and bright sub-mm galaxies (SMGs; \citealt{Smail1997SMG,smail99,frayer20}), these optically faint objects generally have main sequence - like (MS) (e.g. \citealt{daddi10,schreiber15}) star formation rates (SFRs), and are expected to be common in the early universe. Recent studies show that a considerable fraction of these galaxies at high redshifts are missed by the deepest \hst\ surveys due to their extreme faintness in optical and near infrared (NIR). The presence of these $z>2-3$ dusty star-forming population has been 
 demonstrated by a large number of detections with \textit{Spitzer Space Telescope} ($Spitzer$) and \textit{Herschel Space Observatory} ($Herschel$; see \citealt{huang11,caputi12,alcpamp19}), as well as sub-mm and radio regimes \citep{talia21,wang21} specifically by Atacama Large Millimeter/submillimeter Array (ALMA) \citep{simpson14,franco18,yamaguchi19,wang19,williams19,umehata20,caputi21,manning21}. It is yet unclear whether these apparently rare objects form a part of a larger and more elusive population.
\par
There is no clear nomenclature existing for these objects, however the following terms are generally being used interchangeably: $H$-dropouts \citep{wang19}, $K$-faint \citep{smail21}, optically-dark/faint, and \hst-dark/faint galaxies
\citep{franco18,shu22,gomez-guijarro22,xiao22}. 
Among these samples, the well-defined $H$-dropout (or \textit{HST}-dark) populations have been found dominating the cosmic SFR density at the high mass end log$(M_*/M_\odot)>10.5$ at $z>3$ \citep{wang19} but largely missed by previous surveys (e.g. see \citealt{spitler14,wang16}), challenging our understanding of galaxy formation in the early universe.
\par
However, the nature of \hst-dark sources remains to be fully unveiled. Very few spectroscopic redshifts for these objects are available (but see \citealt{swinbank12,williams19,zhou20}) and their rest-frame optical emission is still in the dark, thus complicating the assessment of their physical properties. 
What is their morphology and environment? What role do galaxy mergers play in them? Is the \hst-dark phase a precursor to quenching? To answer these questions, both spectroscopic confirmation and high quality imaging are required to quantitatively analyze their stellar and dust content, morphology, and environment.
\par
Detecting \hst-dark galaxies was only possible with $Spitzer$ Infrared Array Camera (IRAC) imaging, for the stellar emission, or the sub-/mm facilities like ALMA, Northern Extended Millimeter Array (NOEMA) and, to a lesser extent, SCUBA2 (e.g. \citealt{yamaguchi19,gruppioni20,umehata20,manning21,sun22}). Due to the extreme faintness at observed optical/NIR wavelengths, spectroscopically confirming redshifts of these objects is unfeasible with \hst\ and ground-based optical facilities. At the same time, however, (sub)mm interferometry becomes more efficient given the dusty star-forming nature of these sources and the strong negative $K$-correction of dust SEDs. In recent efforts, \citet{jin19,jin22} confirmed redshifts of a sample of 10 optically-dark/faint galaxies at $3.5<z<6$ via CO and [CI] line detections with ALMA and NOEMA, unveiling optically-thick dust in far-infrared (FIR), coupled with high amount of dust attenuation in the optical $3 < A_V <5$. \citet{fudamoto21} discovered two $z\sim7$ \hst-dark galaxies via [CII], which for the first time shed light on the dust-obscured SFR density (SFRD) in the epoch of reionization.
Nevertheless, detailed studies in stellar emission of \hst-dark galaxies have not been feasible until James Webb Space Telescope ($JWST$) became operational. The unparalleled sensitivity, spatial resolution and long wavelength coverage of $JWST$ enable us to unveil the dark nature of this considerable galaxy population for the first time.

\par
Recently, a growing number of near-IR faint galaxies have been identified in early \jwst\ data \citep{barrufet22,carnall22,iani22,labbe22,nelson22}. A few \textit{HST}-dark galaxies were identified as dusty red spiral galaxies at cosmic noon \citep{fudamoto22}. Statistical \textit{JWST} sample from \cite{perezgonzalez22} revealed that 70\% \hst-dark galaxies are massive ($9<\mathrm{log}(M/M_{\odot})<11$) dusty star-forming galaxies at $2<z<6$ with high dust attenuation $2<A_{\rm V}<5$. While these findings are consistent with studies of massive objects performed in the deepest available fields prior to the $JWST$ launch, spatially-resolved studies of \hst-dark galaxies with \jwst\ still have not yet been performed.
\par
In this work, we report the identification, integrated and spatially resolved \textit{JWST} study of a lensed \textit{HST}-dark galaxy and its quiescent companion at $z=2.58$ in the Abell 2744 field.
We adopt a flat $\Lambda$CDM cosmology with $\Omega_{\mathrm{m},0}=0.3$, $\Omega_{\mathrm{\Lambda},0}=0.7$ and H$_0=70$ km s$^{-1}$ Mpc$^{-1}$, a \citet{chabrier} initial mass function, and AB magnitude system.

\section{Observations and Data} \label{sec:obs_data}

\subsection{$JWST$ Imaging Data Reduction}
We homogeneously processed all the publicly available \jwst\ imaging obtained with the Near Infrared Camera (NIRCam) and Near Infrared Imager and Slitless Spectrograph (NIRISS), in the Abell 2744 field. These include the data from: 1) Through the Looking GLASS: A \jwst\ Exploration of Galaxy Formation and Evolution from Cosmic Dawn to Present Day (GLASS; ERS-1324; PI: T. Treu), 2) The \jwst\ UNCOVER Treasury survey: Ultradeep NIRSpec and NIRCam ObserVations before the Epoch of Reionization  (GO-2561; UNCOVER; PIs: I. Labbe, R. Bezanson; \citealt{bezanson22}), 3) DD 2756 (PI: W. Chen).
The images have all been reduced with the \textsc{grizli} pipeline \citep{grizli}. A detailed description of the procedure will be presented in Brammer et al. (in prep.) and \citet{valentino23}.
\par 
We complement our \jwst\ data-set by including all available optical and near-infrared data from the Complete \textit{Hubble} Archive for Galaxy Evolution \citep[CHArGE,][]{kokorev22}. All individual \jwst\ and \hst\ (Hubble Frontier Fields; \citealt{lotz17}) exposures were aligned to the Gaia DR3 \citep{gaia-collaboration2021}, co-added and drizzled \citep{fruchter2002} to a $0.\arcsec02$ pixel scale for the Short Wavelength (SW) NIRCam bands and to $0.\arcsec04$ for all the remaining \jwst\ and \hst\ filters.

\begin{figure*}
\begin{center}
\includegraphics[width=.9\textwidth]{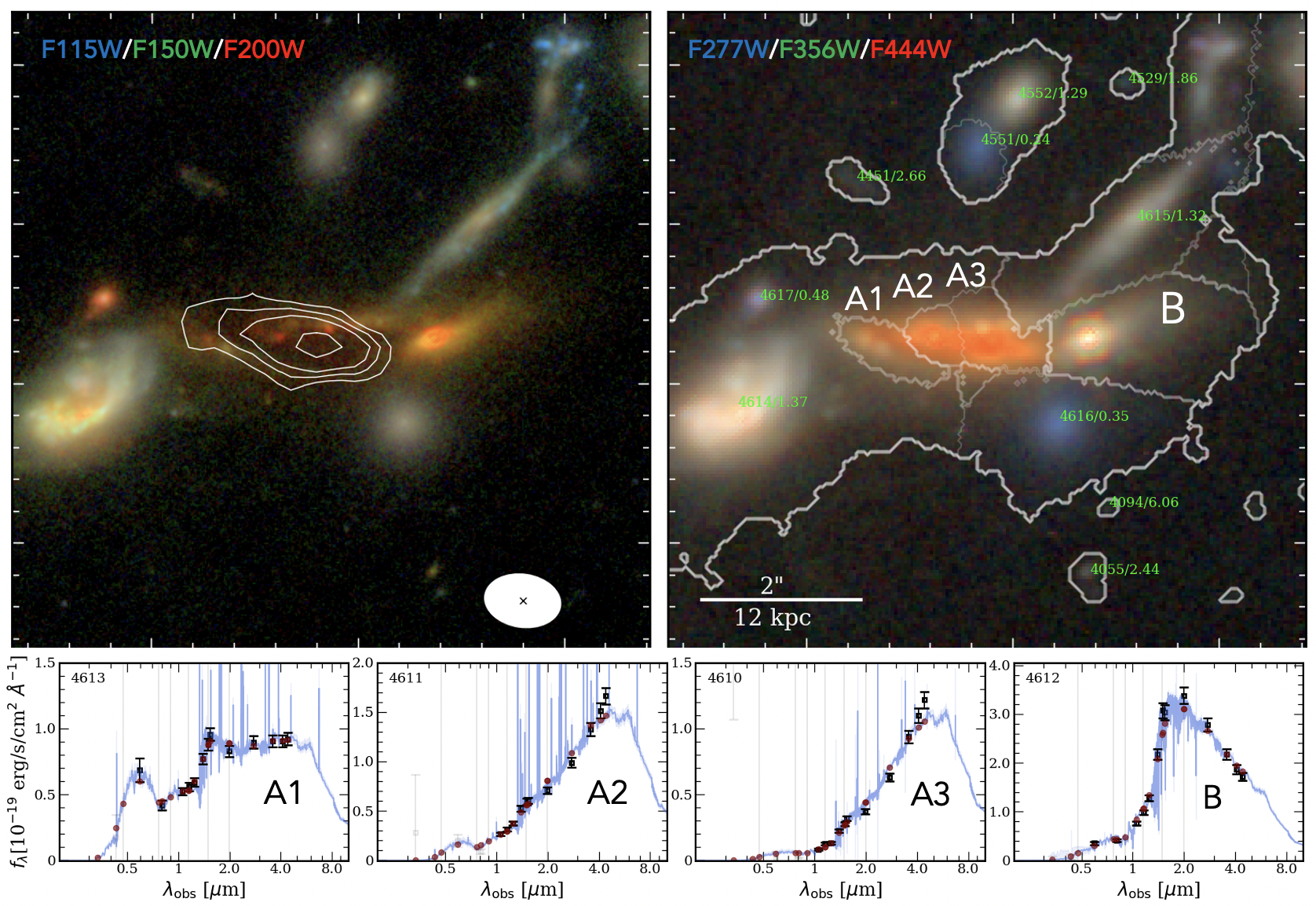}
\caption{\textbf{Top:} Color composite images constructed by using SW (Left) and LW (Right) NIRCam channels. Each cutout is 8\farcs{0}$\times$8\farcs{0}. On the SW stamp, we show the 3, 5, 7 and 12 $\sigma$ contours from the ALMA 1.15\,mm image. The synthesized ALMA beam size is shown at the bottom-right. On the LW stamp, we overlay the segmentation map, in white, and names of our sources of interest. Green labels denote the catalog number and photometric redshifts of neighbouring sources. The physical scale shown corresponds to the image plane. \textbf{Bottom:} Best-fit \textsc{EAZY} SEDs of each component belonging to the dusty spiral (A1, A2 and A3) and the compact quiescent companion (B). Black and grey points show the measured flux density and upper limits, respectively. Red circles represent the template fluxes.}
\label{fig:fig1}
\end{center}
\end{figure*}

\begin{figure*}
\begin{center}
\includegraphics[width=.9\textwidth]{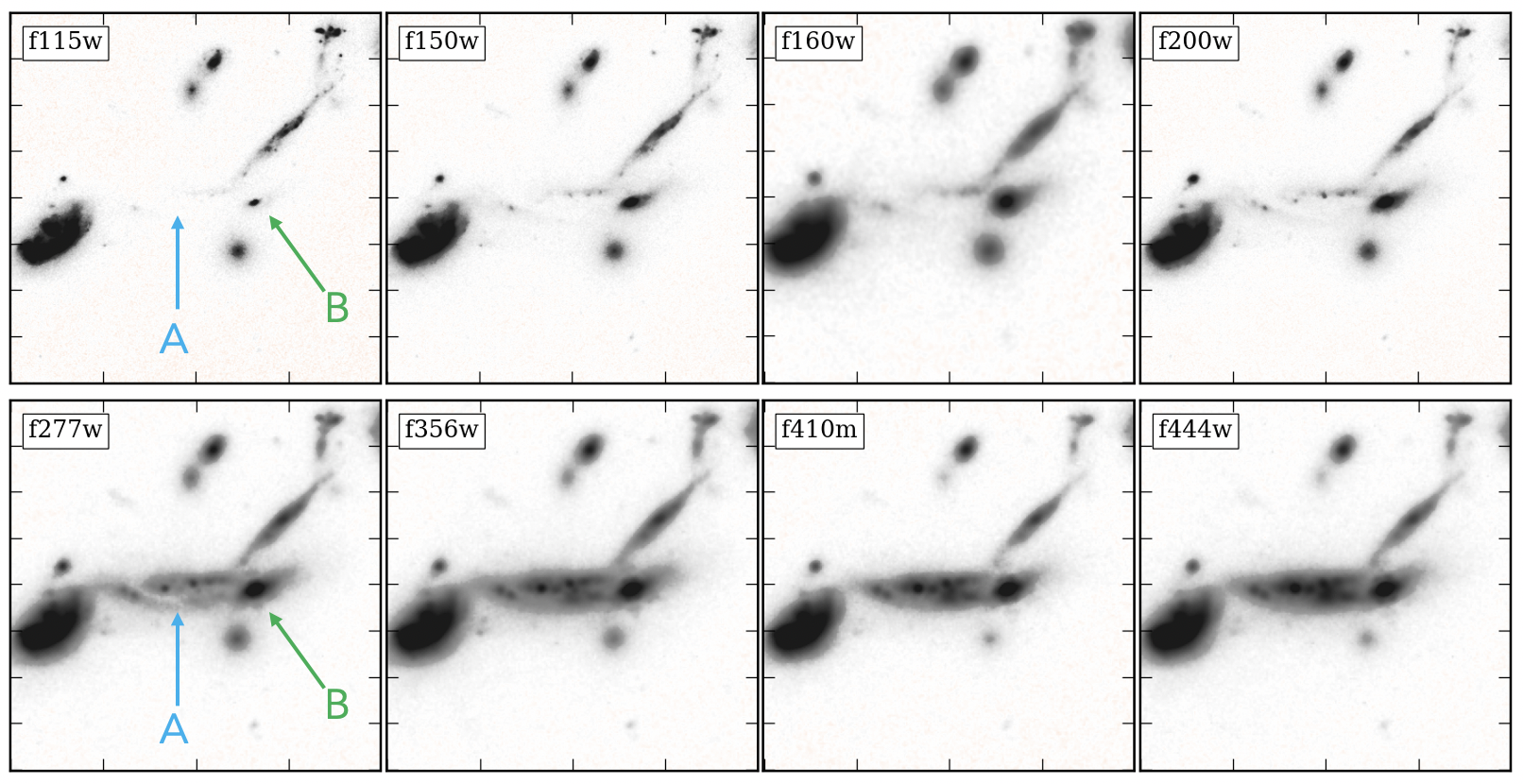}
\includegraphics[width=.9\textwidth]{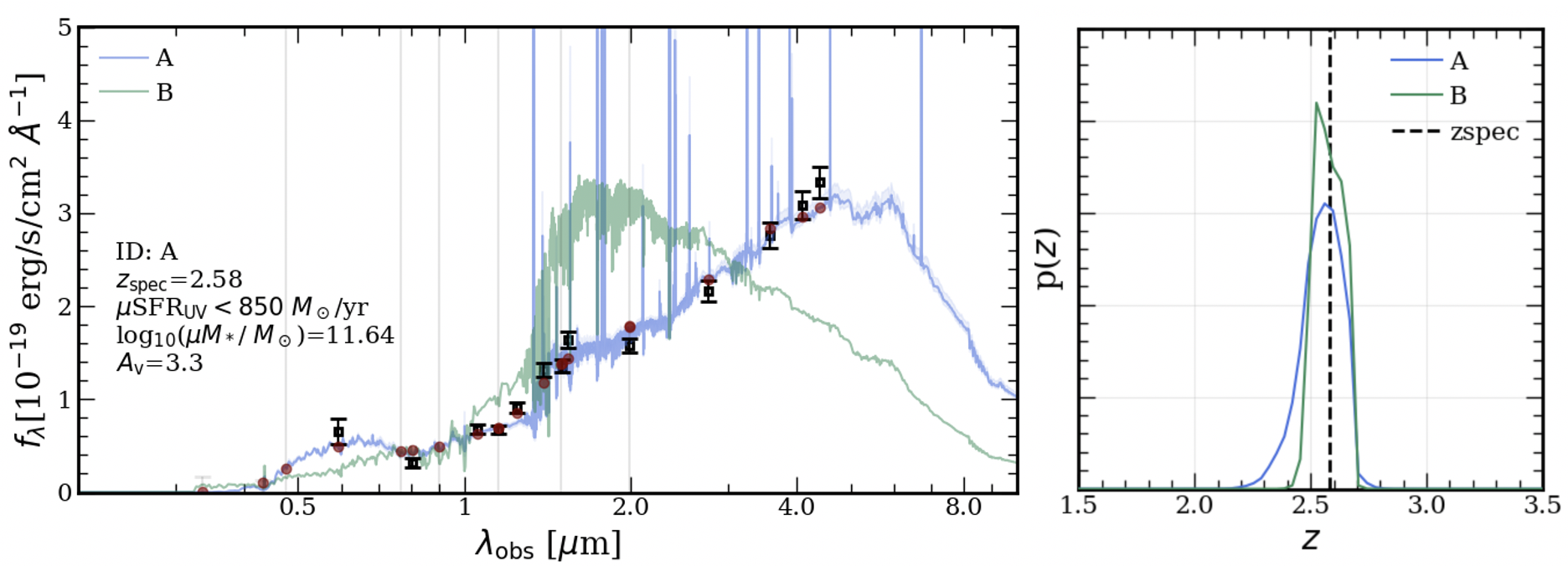}
\caption{\textbf{Top:} Cutouts showing galaxies A and B, in \hst\ $H$-band and all available $JWST$/NIRCam SW and LW observations. Each stamp is 8\farcs{0} per side. 
The galaxy A is largely undetected at $\lambda_{\rm obs}<2$ $\mu$m in both \textit{HST} and \textit{JWST} images, except clumpy sub-structures. A spiral shape becomes apparent in the F277W image and beyond. \textbf{Bottom Left:} Best fit \textsc{EAZY} SED of the entire galaxy A. Blue line show the best fit model, red circles represent the template fluxes, and the black points show the measured flux density. For added clarity we re-plot the SED of galaxy B in light green. \textbf{Bottom Right:} A compilation of the redshift probability distributions ($p(z)$) for the combined source A (blue), and B (light green). The spectroscopic redshift of A, from \citet{sun22} and Bauer et al. (in prep.) is indicated by the dashed black line.
}
\label{fig:fig2}
\end{center}
\end{figure*}

\subsection{Source Extraction}
We extract the sources by using a detection image combined from all noise weighted ``wide'' (W) NIRCam Long Wavelength (LW) filters available, this includes F277W, F356W and F444W. A similar detection method for  \textit{JWST} data was employed in \cite{jin22b}. For source extraction and the segmentation map, we used \textsc{sep} \citep{sep}, a \textsc{Python} version of \textsc{SExtractor} \citep{sextractor}. We extract the 
photometry in circular apertures of increasing size. Correction from the aperture to the ``total'' values was performed by
using the \texttt{flux\_auto} column output by \textsc{sep},
which is equivalent to the \texttt{MAG\_AUTO} from \textsc{SExtractor}, ensuring that for each source only flux belonging to its segment is taken into account. This method was shown to be applicable to both point-like and extended objects as shown in \citet{weaver21}, so we believe it to be adequate for our sources.

We apply an additional correction to account for the missing flux outside the Kron aperture \citep{kron1980}, by utilizing a method similar to the one used in \citet{whitaker11}. In short, this procedure involves computing the fraction of the missing light outside the circularised Kron radius by analyzing curves of growth of the point spread functions (PSF), which were obtained by using the \textsc{WebbPSF} package \citep{perrin14}.  This correction is then applied to the \texttt{flux\_auto} values for each source. For our science, we use the total fluxes, computed from $D$=0\farcs{5} apertures. 

\subsection{Far Infrared Data}
\subsubsection{ALMA}
ALMA Band 6 observations of Abell 2744 were carried out through the ALMA-Frontier Fields survey (PI: F. Bauer) in cycle 3, and the ALMA Lensing Cluster Survey (ALCS; PI: K. Kohno) in cycle 6. The observations reach a rms $\sim 53$ $\mu$Jy  at 1.15 mm at the position of the \hst-dark source. The \hst-dark source has been already identified as an SMG in \citet{laporte17} and \citet{munoz18}, as A2744-ID02. However, no secure optical/NIR identifications were available at that time. In this work, we use the 1.15 mm flux density measurements and a continuum map which will be described in Fujimoto et al. (in prep.). The total flux densities are measured as a peak count in the tapered map [mJy/beam]  after primary beam correction.  

\begin{figure}
\begin{center}
\includegraphics[width=.47\textwidth]{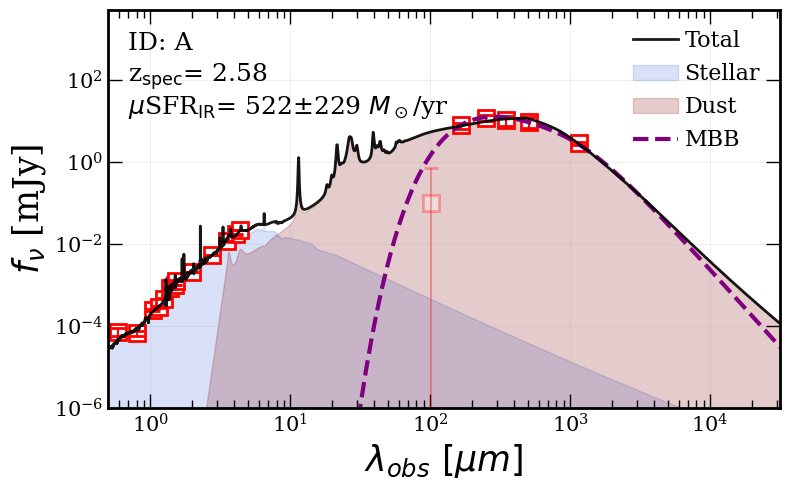}
\caption{Best fit multi-component \textsc{Stardust} fit to galaxy A. The SED is separated into two components - stellar emission (blue), dust (red). The combined emission  (grey line and shaded areas), the observed photometry (red squares) are also displayed. The dashed purple line shown best-fit of the generalized form of the MBB function.}
\label{fig:sed_fits}
\end{center}
\end{figure}

\subsubsection{$Herschel$}
The Abell 2744 cluster was observed by $Herschel$ in the PACS and SPIRE filters as part of the $Herschel$ Lensing Survey (HLS; \citealt{egami10}; E. Egami et al. in prep.).
To complement our SED-fitting analysis, we include the additional
photometric data, spanning a range from $100 - 500$ $\mu$m, from the catalog presented in \citet{sun22}, where the $Herschel$ flux density was deblended and extracted to match the ALMA detections in the field.

\section{Data Analysis}\label{sec:analysis}
\subsection{Segmentation}
From the segmentation image shown in \autoref{fig:fig1}, we visually identify components A1, A2, and A3, which all appear to belong to our main target, the red spiral galaxy. The spiral has been separated into 3 parts due to our choice for the deblending threshold within \textsc{sep}. However, these detection parameters give us the best results for the entire field (see Brammer et al. in prep), and allow us to separate the neighbouring foreground and background objects from our targets. Components A1, A2, and A3 contain the same spiral structure in the images, and align well with the ALMA contours shown in \autoref{fig:fig1}. We therefore 
denote the sum of their flux densities as that corresponding to the entire spiral (galaxy A).
In the following sections, we will refer to galaxy A as both its separate components as well as the sum thereof. In contrast, its companion (galaxy B) appears to be captured by a single segment, with the compact red stellar core in close proximity to galaxy A, and a diffuse yet distinct tail stretching towards the north-west. In addition to the segmentation, in the top panel of \autoref{fig:fig2} we show 8\farcs{0} stamps of both galaxies A and B, as seen by $HST$/F160W and all available $JWST$/NIRCam imaging. 
Finally, we measure the elliptical shapes of both galaxy A and galaxy B by modeling the F444W image with \textsc{GALFIT} \citep{peng02}.

\subsection{EAZY SED Fitting}
To fit the available photometry for our objects, we use the \textsc{Python} version of \textsc{EAZY} \citep{brammer08}. 
In particular, we chose the \textsc{corr\_sfhz\_13} subset of models within \textsc{EAZY}, which contain redshift-dependent SFHs, and dust attenuation. Additionally, we included the best-fit template to the \textit{JWST}-observed extreme emission line galaxy at $z=8.5$ (ID4590) from \citet{carnall22}, which has been re-scaled to match the normalization of the FSPS models. This was done to adequately model potential emission lines with large equivalent widths.
From the best fit \textsc{EAZY} SEDs, we derive photometric redshifts and the stellar population properties which include, but are not limited to the $A_{V}$, $M_*$ and rest-frame colors. 
\par
With \textsc{EAZY} we fit the segments, as well as the full galaxy A, and its compact companion B. The best-fit SEDs to the individual segments are shown in the bottom panel of \autoref{fig:fig1}, while the full combined SED is shown in \autoref{fig:fig2}. We find the best-fit photometric redshifts for all segments, and the companion to be consistent with the available spectroscopic redshift for A, which was obtained from detected $^{12}$CO (3-2) and $^{12}$CO (5-4) lines (\citealt{sun22}, Bauer et al. in prep). The redshift probability distributions - $p(z)$ are shown on the top right panel of \autoref{fig:fig2}. The SED of A1 is quite different to those of A2 and A3, which we believe could be driven by the blue compact structure present within the segment. This blue clump could be a contaminant which belongs to the neighboring object (ID: 4614), or a less dusty star-forming region in the outskirts of galaxy A. 
 Current data, however, does not allow us to provide a definitive answer. Our photo-$z$ solution for A1, however is consistent with the $z_{\rm spec}$.
For all future purposes, we consider galaxy A and galaxy B to be located at the same redshift - $z_{\rm spec} = 2.58$.
\par
The lack of secure detections in rest-frame UV (1250 - 2700 \AA) does not allow \textsc{EAZY} to extract a robust SFR$_{\rm UV}$. Therefore, we manually derive the upper limit on the $L_{\rm UV}$, by considering the median flux density of the filters closest to the rest frame UV. We compute these luminosities at $\lambda_{\rm rest}=1600$ \AA\ and correct for the dust absorption by using the \citet{calzetti00} attenuation law with our best fit $A_{\rm V}$ values. We list our derived properties in \autoref{tab:galaxy_properties}.

\begin{figure*}
\begin{center}
\includegraphics[width=.97\textwidth]{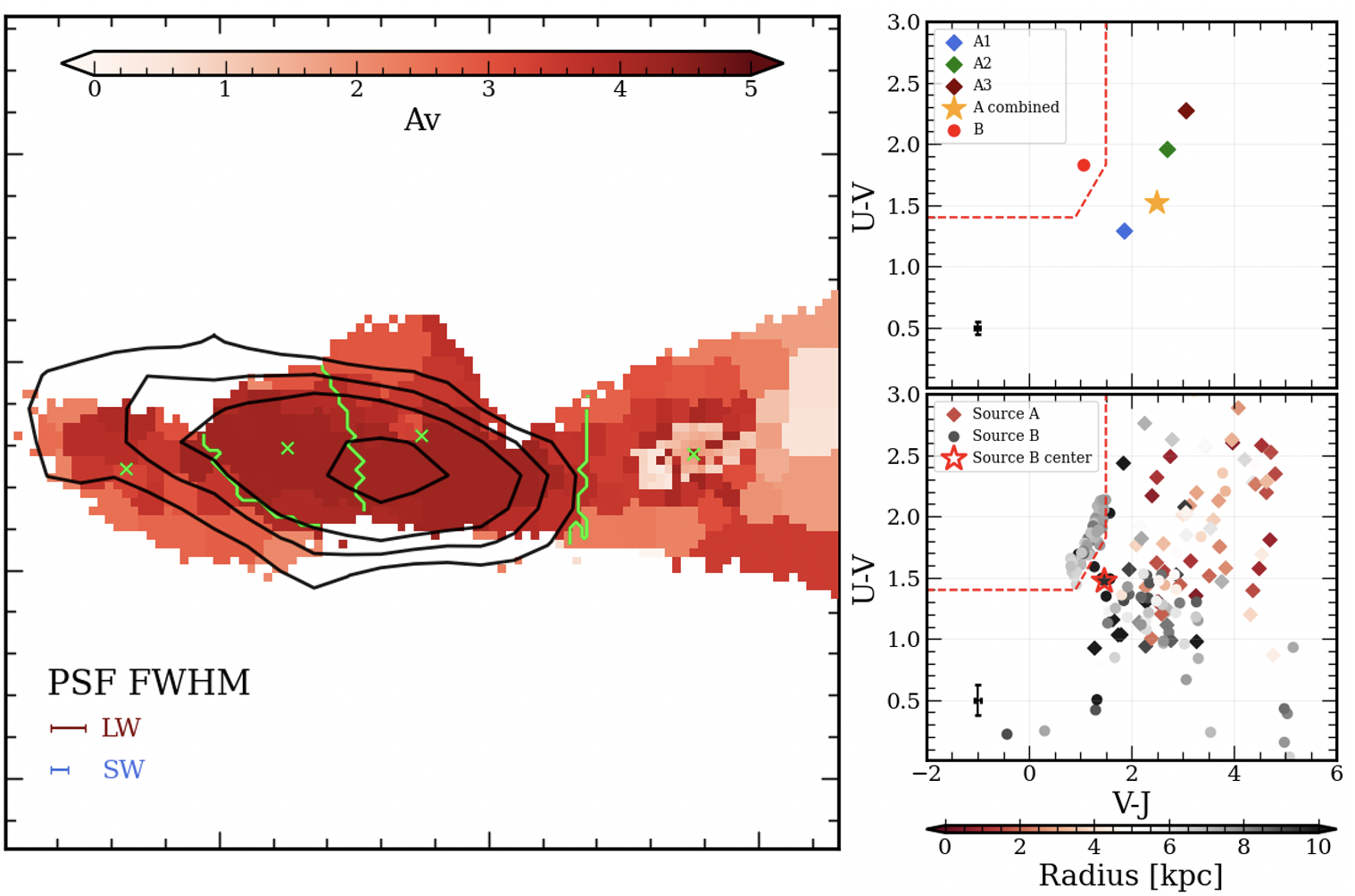}
\caption{\textbf{Left:} Spatially resolved $A_{\rm V}$ map of the 4\farcs{0} stamp centered on galaxy A and B. 
Black contours correspond to the 3, 5, 7 and 12 $\sigma$ levels of the ALMA continuum. In the bottom we also show the average PSF FWHM of SW and LW images. Lime crosses show positions of A1, A2, A3 and B. Solid lime lines highlight the segment borders.
\textbf{Top Right:} Classification of our segments A1,A2, A3, the combined galaxy A and its companion B by using the $U-V$ and $V-J$ rest frame colors. The $UVJ$ galaxy selection separation from \citet{williams09} is shown as a dashed red line. The uncertainty on the colors fits within the marker size. \textbf{Bottom Right:} Spatially resolved $UVJ$ diagram for the 4x4 arcsec$^{2}$ regions around the spiral galaxy. The diamonds denote spatial bins corresponding to galaxy A, and circles belong to galaxy B. The color coding represents the source plane distance (in kpc) from the centroid of galaxy A. The center of galaxy B is highlighted with a red star. A typical uncertainty on the $UVJ$ colors is shown with the error bars on the bottom-left of each figure.}
\label{fig:av_map}
\end{center}
\end{figure*}

\subsection{Fitting of FIR Data}
Out of the two objects, only galaxy A is detected with ALMA, the source also has deblended $Herschel$ flux densities derived in \citet{sun22}.
We perform the SED analysis of combined optical-NIR and FIR photometry with the SED fitting code \textsc{Stardust} \citep{kokorev21}, fixing the redshift to $z_{\rm spec}$.  In short, \textsc{Stardust} relies on a linear combination of optical \citep{brammer08}, AGN \citep{Mullaney2011} and IR dust \citep{dl07,draine14} models, which are fit simultaneously, but without assuming energy balance. We exclude the AGN templates from our fit, as there are no data available in the MIR to reliably constrain the IR AGN emission. From our fit, we derive a range of optical-FIR galaxy properties. \textsc{Stardust} uses the optical templates similar to the ones in \textsc{EAZY}, therefore the stellar population parameters between the two codes are consistent within 0.1 dex. We display the best-fit \textsc{Stardust} SED in \autoref{fig:sed_fits}.
\par
Given its very red colors, it is possible that galaxy A remains optically thick even at FIR wavelengths. As a result, the \citeauthor{dl07} optically thin templates used by \textsc{Stardust} might not be an adequate model for the FIR emission in galaxy A. To alleviate that, we complement our analysis by re-fitting all $\lambda_{\rm rest}>40$ $\mu$m photometry with a generalized form of the modified blackbody (MBB) function (see e.g. \citealt{casey12}). For that we fix the dust emissivity $\beta=1.8$, consistent with results of \citet{magdis2012}, set dust absorption coefficient $\kappa_{\rm 850}$=0.43 m$^2$ kg$^{-1}$ , and let all other parameters vary.
\par
Galaxy B is not detected in ALMA, and presumably by extension neither in $Herschel$, however, we can still use variance of the ALMA map,
to predict the upper limits on the FIR properties. To do that we assume that the FIR SED shapes, and the metallicity for both objects are roughly similar, and then rescale the FIR properties derived for galaxy A, by using the rms of the ALMA map. On the other hand if we chose to use a quenched high-$z$ IR SED template instead, the inferred limit on the $M_{\rm dust}$ would be $\times 2 - 3$ higher, due to the quenched SED being colder. ($T_{\rm dust} \sim20$ K compared to $T_{\rm dust} \sim 35-40$ K; \citealt{magdis21}). Our adopted approach thus gives a conservative estimate of the total dust mass for galaxy B.

\subsection{Spatially Resolved SED Fitting}
We perform spatial binning of the available photometry by utilizing the \textsc{vorbin} \textsc{Python} package \citep{Cappellari2003}. The method uses a two-dimensional spatial binning algorithm by utilising Voronoi tessellation. This splits our image to a given number of spaxels, with a minimum signal-to-noise ratio. To do that we define a 4\farcs{0}-wide square stamp around our area of interest on the detection image (constructed from all LW bands), and set the \texttt{target\_sn} parameter to 200. This threshold was chosen to optimize both the number of bins, and achieve a high SN in the highest amount of available photometric bands. As a result the image was divided into 220 bins, with the median SN of 208. We then use the spatial bins as a segmentation map, and extract flux density for each component from all the available \jwst\ and \hst\ mosaics. From each 4\farcs{0}$\times$4\farcs{0} stamp containing the voronoi bins we additionally subtract the local background computed with \textsc{sep}. 
\par
We fit spatially resolved photometry with \textsc{EAZY}, by adopting the same approach as in the previous section, with the redshift fixed to 2.58. The key parameter which we aim to investigate with the spatially resolved fitting is how much extinction each part of the galaxy, and its companion, have. We show our final $A_{\rm V}$ map in \autoref{fig:av_map}.

\subsection{Magnification Model}
In our work, we use the most recent Abell 2744 mass model presented in \citet{furtak22}. Our galaxies are located significantly far from the caustic lines, and are only marginally affected by gravitational lensing. To produce the magnification map we use the $\kappa$ and $\gamma$ models presented in \citeauthor{furtak22}, with a source redshift $z_{\rm s}$=2.58 for galaxy A, and a cluster redshift $z_{\rm c} = 0.308$. Although our sources are resolved, their magnification varies very little $\Delta\mu\sim0.05$, thus we adopt an effective magnification for the entire system by computing a median $\mu=1.92\pm0.03$ over all segments.

\section{Results} \label{sec:results}

\begin{deluxetable*}{cccccc}
\tablecaption{Derived Galaxy Properties $^\dagger$ \label{tab:galaxy_properties}}
\tablehead{ & A & B & A1 & A2 & A3}
\startdata
\hline
R.A. & 3.5759 & 3.5755 & 3.5764 & 3.5759 & 3.5761\\
Dec & -30.4132 & -30.4132 & -30.4132 & -30.4132 &  -30.4132 \\
$\mu$ & $1.92\pm0.02$ & $1.89\pm0.02$ &  $1.94\pm0.02$ &  $1.92\pm0.02$ &  $1.91\pm0.02$ \\
$r_{\rm eff}$ [kpc] $^1$ & $1.6\pm0.2$ & $0.6\pm0.3$ & - & - & - \\
\hline 
\multicolumn{3}{c}{\textsc{EAZY}: Stellar Population Properties \hfill}\\
\hline
$z_{\rm phot}$ &  2.45 & 2.54 &  2.56 & 2.21 & 2.50 \\
$z_{\rm spec}$ & 2.58 & - &  2.58 & 2.58 & 2.58\\
log$_{10}$($M_*/M_\odot$) &  $11.34 \pm0.05$ & $10.84\pm0.02$ & $10.78\pm0.05$ &  $10.82\pm0.05$ & $10.92\pm0.04$  \\
$A_{\rm V}$ & $3.30\pm0.04$ &  $1.21\pm0.06$ & $2.70\pm0.04$ & $3.52\pm0.05$ & $3.80\pm0.05$ \\
SFR$_{\rm UV}$ [$M_\odot$/yr] & $<450$  & $<4$ & $<206$ & $<208$ & $<40$ \\
SFR$_{\rm UV}$/SFR$_{\rm MS}$ $^{2}$ &  $<3.6$ & $<0.02$ & - & - & - \\
Age [Gyr] & 0.53 & 1.14  & - & - & - \\
\hline 
\multicolumn{2}{c}{\textsc{Stardust}: FIR Properties \hfill}\\
\hline
SFR$_{\rm IR}$ [$M_\odot$/yr] & $290\pm130$ & $<8$ & - & - & - \\
SFR$_{\rm IR}$/SFR$_{\rm MS}$ &  $1.2\pm0.3$ & $<0.06$ & - & - & - \\
log$_{10}$($M_{\rm dust}$/$M_\odot$) & $9.40\pm0.31$& $<8.00$ & - & - & - \\
log$_{10}$($M_{\rm gas}$/$M_\odot$) & $11.27\pm0.32$& $<10.0$ & - & - & - \\
$\delta_{GDR}$$^{3}$ & $76\pm1$& $\sim76$ & - & - & - \\
log$_{10}$ ($M_{\rm dust}/M_*$) & $-2.0\pm0.2$ & $<-2.8$ & -  & - & -  \\
log$_{10}$ ($M_{\rm gas}/M_*$) & $-0.1\pm0.2$ & $<-0.9$ & -  & - & -  \\
\hline 
\multicolumn{2}{c}{Optically Thick MBB \hfill}\\
\hline
$\lambda_0$ [$\mu$m] & $500\pm181$ & - & - & - & -\\
$T_{\rm dust, thick}$ [K] & $60.0\pm4.3$& - & - & - & -\\
$SFR_{\rm MBB}$ [$M_\odot$/yr] & $150\pm80$& - & - & - & -\\
log$_{10}$($M_{\rm dust}$/$M_\odot$) & $8.67\pm0.05$& - & - & - & -\\
log$_{10}$ ($M_{\rm dust}/M_*$) & $-2.6\pm0.3$ & - & -  & - & -  \\
\enddata
\begin{tablenotes}
\item[$\dagger$] \footnotesize{These properties are corrected for the lensing magnification}. \\
\item[1] \footnotesize{Measured on the F444W image.}
\item[2] \footnotesize{Using MS relation from \citet{schreiber15}. \hfill\\
\item[3] \footnotesize{For galaxy B we assume the same metallicity as for A, given they similar stellar mass.\hfill} \\
\hfill}
\end{tablenotes}
\end{deluxetable*}

\begin{figure*}
\begin{center}
\includegraphics[width=.9\textwidth]{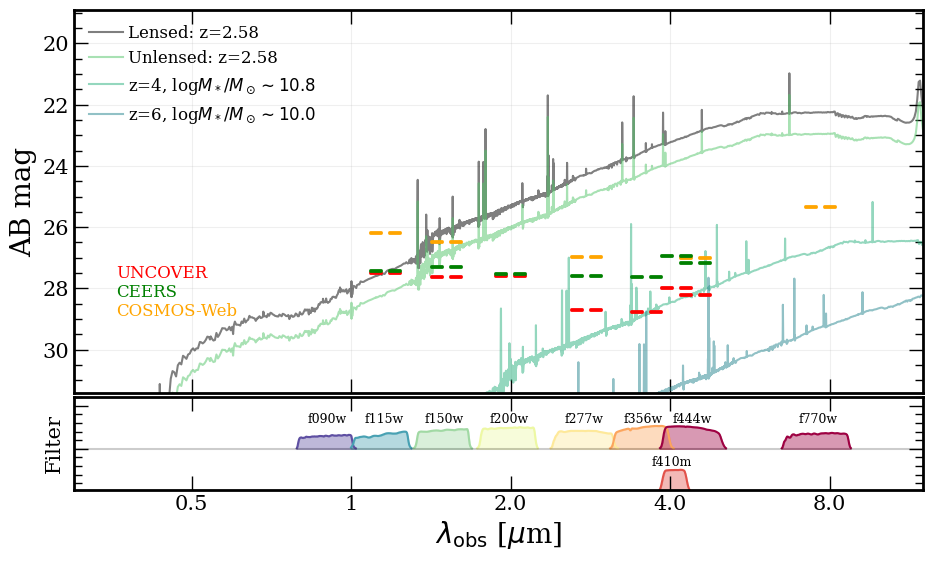}
\caption{SED of the $HST$-dark galaxy A in the unlensed and higher-$z$ cases.  Overlaid are NIRCam and MIRI detection limits of major surveys \citep{bagley22,bezanson22,casey22}. Quoted detection limits are $5\sigma$ depth for extended emission within a $D$=0\farcs{5} aperture. At the bottom of the figure we show the transmission curves of the relevant NIRCam and MIRI filters.}
\label{fig:det_feas}
\end{center}
\end{figure*}

Using \jwst/NIRCam images in the Abell 2744 lensed cluster field, we identify a merging system which consists of an extremely dusty, \hst-dark, star-forming spiral galaxy (galaxy A, R.A. 3.5759 Dec: -30.4132) and a compact quiescent companion (galaxy B). Galaxy A was initially identified as an \hst-dark SMG in \citet{laporte17}, based on its clear detection in ALMA and $Spitzer$/IRAC, and has been spectroscopically confirmed at $z_{\rm spec}=2.58$ via the detection of $^{12}$CO (3-2) and $^{12}$CO (5-4) lines (\citealt{sun22}, Bauer et al. in prep). 
We list the derived physical parameters in \autoref{tab:galaxy_properties} and describe the results as following.

\subsection{\textit{JWST} View of \hst-dark Galaxy A}

As shown in \autoref{fig:fig1} and \autoref{fig:fig2}, the galaxy A is largely invisible in \textit{HST} images, and only two diffuse substructures are detected in F160W ( depth $\sim$27 mag). Strikingly,  it also drops out in \jwst\ F115W image with depth $\sim28.5$ mag. The two diffuse structures in \textit{HST}/F160W are resolved into compact star-forming clumps in \jwst\ F150W and F200W images. 
In contrast, galaxy A is well detected in all LW channels, presenting a nearly edge-on disk morphology. Remarkably, a spiral structure in galaxy A is well uncovered in F277W, as shown by the arrow in \autoref{fig:fig2}. These together indicate that the \hst-dark galaxy is a nearly edge-on spiral galaxy, suggesting high dust thickness given its extreme faintness in \textit{HST} and \textit{JWST} SW bands.

SED fitting in \autoref{fig:fig2} shows that galaxy A is extremely dust-obscured with an average $A_{\rm V}=3.30\pm0.04$ mag.
The $A_{\rm V}$ is comparable with the high attenuation in other \hst-dark or optically-faint/dark populations \citep{wang19,smail21,jin22,xiao22}, and is highly robust thanks to the state-of-the-art \jwst\ photometry.
The dust extinction gradient is $\Delta A_{\rm V} \sim 1$, with the $A_{\rm V}$ increasing from component A1 to component A3. The upper limits on SFR$_{\rm UV}$ appear to decrease as we move to component A3, which lies closest to galaxy B.  SED fitting to both optical and FIR photometry place galaxy A within the scatter of the main sequence of star formation \citep{schreiber15} according to both the rest-frame UV, and IR-derived SFRs. Further investigation of the rest-frame colors, presented in the right panel of \autoref{fig:av_map}, confirms that galaxy A, as well as its constituents A1, A2 and A3 are located in the dusty star-forming part of the $UVJ$ diagram \citep{williams09}.
\par
Using the spatially resolved photometric analysis shown in \autoref{fig:av_map}, we find the $A_{\rm V}$ to be the largest in the middle, coincident with the peak of ALMA emission, reaching $A_{\rm V}\approx4.5$, and weaker towards the outskirts where $A_{\rm V}\approx3$. 
These extreme values of dust extinction are consistent with similar \hst-dark populations presented in \citet{jin22}. In our spatial analysis we also infer a dearth of dust obscuration in the north and south-east, compared to the rest of the galaxy. This gradient of $A_{\rm V}$ is seen both in the analysis of larger components A1, A2 and A3 as shown in the result of our \textsc{EAZY} SED fitting \autoref{tab:galaxy_properties}, and the spatially resolved analysis in \autoref{fig:av_map}. This explains why previous observations with \hst\ were only able to identify a small portion of this galaxy. We also note that the obscured stellar disk seen in components A1, A2 and A3 traces well the ALMA continuum emission at 1.1 mm. In fact, when corrected for lensing effects we find an effective radius in F444W of $r_{\rm eff}\sim1.6$ kpc, which is consistent with the 1.1 mm size calculated in \citet{laporte17}.
\par
It is important to point out that a combination of 
$A_{\rm V}\sim4.5$ and optically thick FIR might imply that a bulk of on-going star formation and potentially significant amounts of $M_*$ could be obscured up to $A_{\rm V}\sim1000$ (see \citealt{simpson17}). This could 
mean that even rest-frame NIR observations are not sufficient when dealing with highly obscured systems.
\par
In \autoref{fig:sed_fits}, we show the NIR+FIR SED of galaxy A. Both the optically thin \textsc{Stardust} and the optically thick MBB fit well the $Herschel$/SPIRE and ALMA photometry with consistent SFR$_{\rm IR}\sim300~M_\odot$ yr$^{-1}$, while the optically thick MBB provides a better fit to the upper limits in $Herschel$/PACS.
The resultant SFR$_{\rm IR}$ is broadly consistent with the UV-based upper limit. 
Notably, the MBB fitting confirms the optically thick dust in this galaxy, where the dust emission is optically thick up to $500\pm181$ $\mu$m rest frame. 
This results in $M_{\rm dust}$ that is $\sim3\times$ lower than that from optically thin fitting and a warm dust temperature of $T_{\rm dust}\sim$60\,K, which are consistent with the results in \citep{cortzen20,jin22}. The dust to stellar mass ratio of galaxy A ($M_{\rm dust}/M_*$ $\sim 10^{-2.6}$) is lower than a typical value for MS galaxies at that mass and redshift ($M_{\rm dust}/M_*$ $\sim 10^{-2.0}$) as show in the literature \citep{scov17,tacconi18,kokorev21,liu21}, but still within the scatter, supporting the idea that it is a typical dusty star-forming galaxy.

The judgement whether the optically thin \citeauthor{dl07} models or the generalized form of the MBB function fit our data best can only rely on the $\chi^{2}$ of both fits, which are broadly consistent. Further ALMA observations at high spatial resolution, combined with spatially resolved $JWST$ photometry would therefore be required to definitively ascertain which physical model best describes the FIR dust in galaxy A.

\subsection{Quiescent Companion Galaxy B}
In the vicinity of the dusty spiral, a compact red object, galaxy B, was identified at the similar redshift of the spiral. 
The $p(z)$ solution for galaxy B matches well with the $z_{\rm spec}$ of galaxy A, as shown in the right panel of \autoref{fig:fig2}, suggesting that both belong to the same system.
The \textsc{GALFIT} derived de-lensed effective radius for galaxy B was calculated to be $r_{\rm eff}\sim 0.6$\,kpc. We compute the projected physical distance between the centers of the two galaxies to be 9 kpc, corrected for the lensing effects. The close proximity of the two galaxies suggests that a potential merging event is currently taking place. By examining the cutouts in \autoref{fig:fig1} and \autoref{fig:fig2} we note the existence of a diffuse tail to the upper-right part of the object. Provided that two galaxies are indeed merging, the streamers could be caused by the tidal interaction leading to the spread of the far side of the quiescent galaxy as a tidal arm. In addition to that, the region of galaxy A which is the closest to the compact companion also appears to be "bowed" and disturbed. Both of these features further consolidate our merger scenario, presumably near first passage given that both galaxies are still largely intact. 

In the $UVJ$ diagram (\autoref{fig:av_map}), galaxy B is located in the quiescent regime, which is consistent with the low SFR limits in \autoref{tab:galaxy_properties}. The spatially-resolved diagnostics (\autoref{fig:av_map}-bottom right) also indicate that most components in galaxy B are quiescent, further supporting its quiescent nature. 
Interestingly, the center of galaxy B drops out from the quiescent regime in the resolved $UVJ$ diagram (\autoref{fig:av_map} - right), and it is point-like in LW color image (\autoref{fig:fig1}). 
As an additional verification, we fit optical quasar \citep{shen16} and Phoenix star \citep{brammer08} models to galaxy B photometry and found no adequate solutions, ruling out both the quasar and stellar templates. Therefore, these together might suggest that it hosts moderate AGN activity.

\par
The derived upper limit on the UV ($<4$ $M_\odot$/yr) and IR SFRs ($<8$ $M_\odot$/yr), in conjunction with the position of galaxy B on the $UVJ$ diagram securely place it in the quiescent regime. We find that galaxy B is $\sim3\times$ lesser massive in stellar mass than the red spiral, and significantly less dusty as indicated by the smaller average $A_{\rm V} \sim 1$ and non-detection in ALMA.
Using the upper limit from ALMA, we derive a dust mass log$_{10}$$(M_{\rm dust}/M_\odot)<8.0$ and a dust fraction log$_{10}$$(M_{\rm dust}/M_{*})<-2.8$, which are consistent with $z>2$ quiescent samples in literature \citep{gobat18,magdis21,whitaker21}.

\subsection{Large Scale Environment}

The deep \jwst\ data-sets and large photometric catalog enable us to quantify the environment in an unbiased manner.
In order to assess the environment of this galaxy pair, we conducted overdensity analysis following the method adopted in \cite{Sillassen2022}. We pre-selected 527 sources with $2.4<z_{\rm phot}<2.8$ in the \textit{JWST} photo-$z$ catalog of Abell 2744 produced with \textsc{grizli} and \textsc{EAZY}, and mapped the distance of the fifth and tenth neighborhood $\Sigma_{\rm 5th}$ and $\Sigma_{\rm 10th}$ for each source. 
The average galaxy densities are $\Sigma_{\rm 5th}=30$~arcmin$^{-2}$ and $\Sigma_{\rm 10th}=22$~arcmin$^{-2}$, respectively.
We find that the  $\Sigma_{\rm 5th}$  and $\Sigma_{\rm 10th}$ of the $HST$-dark galaxy are 23 and 33~arcmin$^{-2}$  respectively, which both have less than 1$\sigma$ significance comparing to the average in this field. Therefore, there is almost no overdensity of galaxies around galaxies A and B, and their environment is as common as in random fields.

\section{Discussion} \label{sec:discussion}
\subsection{Detectability at Higher-$z$}
The \textit{HST}-dark galaxy A is unambiguously detected and resolved with \jwst\, though, an important question which remains is whether \jwst\ can detect (see a similar discussion for $HST$ in \citealt{dey99}) similarly dusty objects that are unlensed at higher redshift, especially the \jwst\ prospects on dusty galaxies in the Epoch of Reionization ($z>6$).
To answer this, we compare the best-fit SED model of galaxy A at different redshifts with the depths of major \textit{JWST} surveys, including UNCOVER (PIs: I. Labbe, R.Bezanson; \citealt{bezanson22}), CEERS (PI: S. Finkelstein; \citealt{bagley22}) and COSMOS-Web (PIs: J. Kartaltepe, C. Casey; \citealt{casey22}). 
At higher redshift, galaxies are less massive and more compact given the mass-size relation (e.g. see \citealt{vanderwel14}).
Therefore, we scale the intrinsic SED accounting for multiple effects: (1) surface brightness dimming with $(1+z)^4$; (2) using the merger trees in \cite{jin22b}, we assumed that the galaxy has log$(M_*/M_\odot)=10.8$ at $z=4$ and log$(M_*/M_\odot)=10$ at $z=6$; (3) adopting the stellar mass-size relation at different redshift from \cite{Costantin2022} for simulated $3<z<6$ galaxies in \textit{JWST} F356W image, i.e., $r_{\rm eff}\propto(1+z)^{-0.14}$; (4) accounting for the physical scale to angular scale scaling at the assumed redshift.
In order to consistently compare to the depths of \textit{JWST} surveys. We normalized the SED to extended emission in an aperture of 0\farcs{5} diameter. The tabulated  $5\sigma$ detection limits of the \jwst\ surveys are mostly for point-like sources thus, where appropriate, we normalized them to extended emission depths by converting the total flux densities back to the one measured within the same aperture using the empirical PSF curves of growth. 

We show the scaled SEDs alongside the detection limits in \autoref{fig:det_feas}. Our results indicate that all the major \jwst\ surveys considered are able to detect comparably dusty and unlensed objects at $z\sim2.6$ at $\lambda\geq1.5\mu$m.
However,  at higher redshift the picture changes dramatically, mainly due to the resolved surface brightness dimming with $(1+z)^4$. As shown in \autoref{fig:det_feas}, at $z\sim4$ analog galaxies would drop out in all SW images and become "3 or 4 $\mu$m-only" objects in the deep UNCOVER survey. As concluded by \citet{Zavala2022,Naidu2022}, these 2 $\mu$m dropouts could masquerade as the $z>15$ candidates and have a severe impact on studies in the most distant universe.
Strikingly, at $z\sim6$ similarly dusty objects would be completely invisible in any \jwst\ bands, and become \jwst-dark. Even considering the extreme case with log$(M_*/M_\odot)=11$, such massive stellar mass at $z=6$ is approaching the most massive $M_*$ allowed by $\Lambda$CDM cosmology \citep{behroozi18} and only scales up the SED by a factor of 10, however the SED would be still below the detection limits.
Given that massive star-bursting SMGs have been found at $z\sim4-7$ \citep{capak11,walter12,riechers13,zavala17,marrone18,casey19,fudamoto21,jin19,endsley22,jin22}, and \hst-dark sources are suspected to be dominant at $z=3-6$ \citep{wang19,shu22}, dusty disk galaxies on the MS are thus expected to be more common at that epoch (e.g., \citealt{neeleman20}), however a majority of these objects would be \jwst-dark. Therefore, identification of extremely dusty MS galaxies in the epoch of reionization is challenging with \jwst.
We note that $HST$-dark galaxies at high-$z$ might not have similarly extreme attenuation as this source. Further studies are therefore required to understand whether the dust attenuation of optically faint galaxies is lower at high-$z$, and how likely it is for \jwst\ to detect these types of objects.

\subsection{Quiescence and Extreme Obscuration in a Pair --- Another Jekyll \& Hyde}
The \jwst\ imaging reveals galaxy A, previously classified as \hst-dark, to be a dusty star-forming disk. Its companion, a compact galaxy B, appears to be completely quenched, as seen from its SFR$_{\rm UV}$, $UVJ$ colors, and lack of IR dust emission while being almost as massive as galaxy A. A similar pair ZF-COSMOS-20115 at $z=3.7$, consisting of a quiescent galaxy and an optically dark star-forming galaxy, was reported by \citet{schreiber18}, who dubbed them 'Jekyll' and 'Hyde', respectively. Similar to galaxy A, Hyde is extremely dust obscured with $A_{\rm V}\sim3.5$, and is only seen in IRAC and ALMA. The [CII] line in Hyde shows a clear rotating disk profile. After correcting for the lensing magnification, galaxies A and B are similarly massive to Jekyll and Hyde in stellar masses. Where \citeauthor{schreiber18} report masses of log$_{10}(M_*/M_\odot)\sim10.9$ and log$_{10}(M_*/M_\odot)\sim11.1$ for Hyde and Jekyll respectively, we find 
log$_{10}(M_*/M_\odot)\sim11.3$ and log$_{10}(M_*/M_\odot)\sim10.8$  for galaxies A and B. While the quiescent Jekyll was found to be roughly 3 times more massive than Hyde, we find exactly the opposite in our case.
Moreover, the dusty disk of Galaxy A appears to be more star-forming, at $\sim 1.2 \times$ above the MS, compared to a moderately suppressed mode of star formation in Hyde, which \citeauthor{schreiber18} hypothesize stopped forming stars $\sim 0.2$ Gyr prior. On the other hand, galaxy B has comparably low SFR to Jekyll with SFR$<4$ $M_\odot$/yr. The de-magnified size of galaxy B is also consistent with the stellar size of Jekyll within $1\sigma$. In addition to that, we infer that galaxy A has a 5$\times$ larger gas mass reservoir compared to Hyde, as derived from the $M_{\rm dust}$, where we used a solar metallicity-like $\delta_{\rm GDR}$ akin to \citeauthor{schreiber18}. However, the differences in the SED modelling, and more specifically the fact that we use graphite-based dust models \citep{draine07}, as opposed to the amorphous carbon grain templates used in \citeauthor{schreiber18}, might result in a factor 2.6 larger dust, and therefore gas mass. Similar to the quiescent Jekyll, no sign of significant star formation is seen in galaxy B.
\par
Given the similarity of the two systems, one could suspect if Jekyll \& Hyde would evolve to the galaxy pair in this work.
We note that Jekyll \& Hyde are at $z=3.7$ (0.8 Gyr earlier than at $z=2.58$). This time is sufficient for the dusty Hyde to deplete its molecular gas and become a quiescent galaxy at $z=2.5$, similar to galaxy B. While for the already quenched Jekyll, its star formation is unlikely to be rejuvenated to match that of galaxy A. Therefore, Jekyll and Hyde are more likely to be evolving into a quiescent pair rather than this system.

\section{Conclusions} \label{sec:conclusions}
Using the state-of-the-art \jwst\ imaging and FIR data, we present the first spatially resolved analysis of a gravitationally lensed \hst-dark galaxy at $z=2.58$. Our conclusions are as follows:
\\
[2mm]
$\bullet$  \jwst\ imaging unveiled the nature of this \hst-dark galaxy at $z=2.58$ as a nearly edge-on dusty spiral with log$_{10}( M_*/M_\odot)\sim11.3$, and identified it as part of a merging system with a massive quiescent companion. The quiescent companion, located at a lens-corrected physical separation of $\approx$ 9\,kpc from the spiral, has an intrinsic stellar mass log$_{10}(M_*/M_\odot)\sim10.8$ and shows potential AGN activity.
\\
[2mm]
$\bullet$ High resolution \jwst\ imaging shows a diffuse structure trailing behind the quiescent companion, implying potential tidal interaction. This further implies the merger nature of this system. The stellar disk in galaxy A, appears to be relatively undisturbed, which could imply that the merger is indeed near first passage. 
\\
[2mm]
$\bullet$ The integrated NIR and FIR SEDs of the dusty spiral (galaxy A) exhibit high attenuation in both the optical and FIR bands ($A_V\gtrsim3$, $\lambda_0\sim500$ $\mu$m), consistent with \hst-dark or optically-dark/faint populations.
\\
[2mm]
$\bullet$  Using spatially resolved SED fitting, we find that the high attenuation $A_V$ is largely flat across the galaxy A ($A_{\rm V}\sim$4 over 57 kpc$^2$), and decreases towards the outskirts. We also find that the high $A_V$ region spatially matches to the ALMA 1.15 mm continuum emission.
\\
[2mm]
$\bullet$ Detecting unlensed analogs of dusty \textit{HST}-dark objects at high-$z$ will be complicated by rapidly dimming surface brightness, and extreme amount of dust obscuration, such that similar galaxies like this will be completely \jwst-dark in the Epoch of Reionization ($z>6$).
\\
[1mm]
Spectroscopic followups with NIRSpec IFU would be ideal to detect optical lines given this system fits perfectly the FoV ($\sim 3\farcs{0}$), which would further reveal the stellar population, merger dynamics, AGN activity, as well as quenching mechanisms. Robust measurements of molecular gas content and dynamics can be achieved by observing the [CI] with ALMA \citep{bothwell17,valentino20} and the mid-IR H$_2$ lines with MIRI, which will enable us to better infer evolution paths of \textit{HST}-dark and quiescent galaxies.
\acknowledgments
We thank the anonymous referee for a number of constructive suggestions, which helped to improve this manuscript.  We thank Ian Smail and Fengwu Sun for their helpful suggestions. VK and KIC acknowledge funding from the Dutch Research Council (NWO) through the award of the Vici Grant VI.C.212.036. SJ acknowledges the financial support from the European Union's Horizon research and innovation program under the Marie Sk\l{}odowska-Curie grant agreement No. 101060888. GEM acknowledge financial support from the Villum Young Investigator grant 37440 and 13160 and the The Cosmic Dawn Center (DAWN),funded by the Danish National Research Foundation under grant No. 140. PD \& MT acknowledge support from the NWO grant 016.VIDI.189.162 (``ODIN"). PD also acknowledges support from the European Commission's and University of Groningen CO-FUND Rosalind Franklin program. FEB acknowledges support from ANID-Chile BASAL CATA FB210003, FONDECYT Regular 1200495 and 1190818, and Millennium Science Initiative Program  – ICN12\_009.  
KK acknowledges the support by JSPS KAKENHI Grant Number JP17H06130 and the NAOJ ALMA Scientific Research Grant Number 2017-06B. This work is based on observations made with the NASA/ESA/CSA James Webb Space Telescope. The data were obtained from the Mikulski Archive for Space Telescopes at the Space Telescope Science Institute, which is operated by the Association of Universities for Research in Astronomy, Inc., under NASA contract NAS 5-03127 for JWST. All the $JWST$ and $HST$ data used in this paper can be found in MAST: \dataset[10.17909/ajen-7n42]{http://dx.doi.org/10.17909/ajen-7n42}. This paper makes use of the following ALMA data: ADS/JAO.ALMA\#2018.1.00035.L and 2013.1.00999.S. ALMA is a partnership of ESO (representing its member states), NSF (USA), and NINS (Japan) together with NRC (Canada), NSC and ASIAA (Taiwan), and KASI (Republic of Korea) in cooperation with the Republic of Chile. The Joint ALMA Observatory is operated by ESO, AUI/NRAO, and NAOJ.

\software{Astrodrizzle \citep{astrodrizzle}, EAZY \citep{brammer08}, FSPS \citep{conroy09}, GALFIT \citep{peng02}, grizli \citep{grizli}, sep \citep{sep}, SExtractor \citep{sextractor}, Stardust \citep{kokorev21}, vorbin \citep{Cappellari2003}.}

\facilities{\jwst, \hst, $Herschel$, ALMA}
\clearpage

\bibliographystyle{aasjournal}
\bibliography{refs}

\end{document}